\documentclass[final,english]{elsarticle}
\usepackage[T1]{fontenc}
\usepackage[latin9]{inputenc}
\usepackage{varioref}
\usepackage{units}
\usepackage{amsmath}
\usepackage{amssymb}
\usepackage{graphicx}
\usepackage{esint}
\usepackage{caption}
\usepackage{subcaption}

\pdfpageheight\paperheight
\pdfpagewidth\paperwidth

\journal{Elsevier}
\usepackage{upgreek}
\usepackage[bookmarks,bookmarksopen,bookmarksdepth=6]{hyperref}
\usepackage{cases}
\usepackage{url}

\usepackage{a4wide}

\usepackage{newtxtext,newtxmath,amsmath}
\makeatother

\usepackage{babel}
\begin{document}

\title{Electro-optical imaging of electric fields in silicon sensors}

\author[]{R.~Klanner }
\author[]{A.~Vauth \corref{cor1}}

\cortext[cor1]{Corresponding author. Email address: Annika.Vauth@desy.de, Tel.: +49 40 8998 4728}
 \address{ Institute for Experimental Physics, University of Hamburg,
  \\Luruper Chaussee 147, D\,22761, Hamburg, Germany.}



\begin{abstract}

 A conceptual set-up for measuring the electric field in silicon detectors by electro-optical imaging is proposed.
 It is based on the Franz--Keldysh effect which describes the electric field dependence of the absorption of light with an energy close to the band gap in semiconductors.
 Using published data for silicon, a measurement accuracy of 1 to 4\,kV/cm is estimated.
 The set-up is intended for determining the electric field in radiation-damaged silicon detectors as a function of irradiation fluence and particle type, temperature and bias voltage.
 The overall concept and the individual components of the set-up are presented.

\end{abstract}


\maketitle
 \pagenumbering{arabic}

\section{Introduction}
 \label{sect:Introduction}

 Segmented silicon detectors for precision tracking are the central detectors of most high-energy physics experiments.
 One of the major limitations of silicon sensors is radiation damage, in particular bulk radiation damage from energetic particles.
 In order to understand radiation-damaged silicon sensors, simulate their performance and optimize their design, the knowledge of the electric field in the sensor is important.

 So far no satisfactory method for determining the electrical field in radiation-damaged silicon sensors is available.
 The methods Edge-TCT (Transient Current Technique)~\cite{Kramberger:2010} and TCT-TPA (TCT-Two Photon Absorption)~\cite{Fernandez:2017} provide precise information on the charge collection efficiency, however only an approximate estimate of the electric field~\cite{Klanner:2020}.
 One reason is that the information on the electric field is obtained from the initial sub-nanosecond rise of the pulse, which is strongly influenced by the time response of sensor and readout.
 In principle TCAD simulations can be used to estimate the electric field.
 However, the big number of different defects and defect clusters present a major problem, and so far no satisfactory description of all experimental results on radiation-damaged silicon sensors has been achieved~\cite{Schwandt:2018}.

 The method proposed aims for a direct imaging of the electrical field.
 It is based on the theoretically predicted~\cite{Franz:1958, Keldysh:1958} and experimentally observed \cite{Wendland:1965} electric field dependence of the absorption of photons with energies close to the silicon band gap (1.124\,eV at 300\,K).
 As a first step it is planned to apply the method to non-irradiated sensors, where the electric field is precisely known, and measure precisely the Franz--Keldysh effect.
 If successful, the method will then be used to study the electric field in irradiated sensors as a function of bias voltage, temperature, irradiation fluence and particle type.
 In~\cite{Scharf:2020} it has been observed that the states in the band gap produced by radiation damage change the light absorption, which could be described as a reduction of the silicon band gap.
 With the proposed set-up, the temperature and radiation-damage dependence of this effect can be determined using pieces of silicon irradiated by different particles to different fluences.

 The idea of electro-optical imaging of solid-state detectors is not new.
 In~\cite{Broer:2010} an extensive presentation of electro-optical imaging for CdS is given; however, nothing is said about silicon.

 \section{Measurement principle and set-up}
  \label{sect:Principle}

\begin{figure}[!ht]
   \centering
   \begin{subfigure}[b]{0.456\textwidth}
     \includegraphics[width=\linewidth]{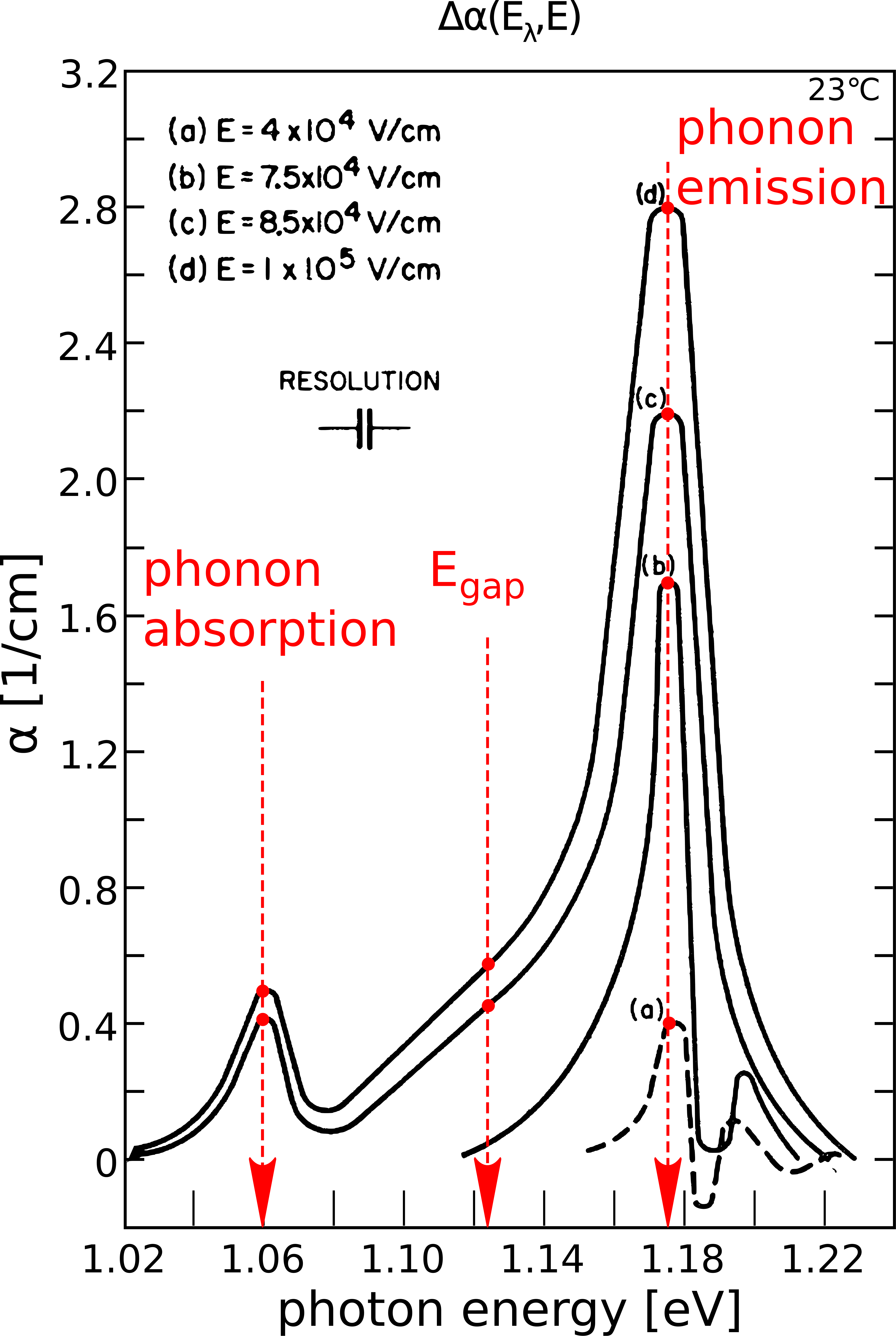}
     \caption{ }\label{subfig:Wendland}
   \end{subfigure}
   \hfill
   \begin{minipage}[b]{0.512\textwidth}
     \begin{subfigure}[t]{\linewidth}
       \hfill\includegraphics[width=0.85\linewidth]{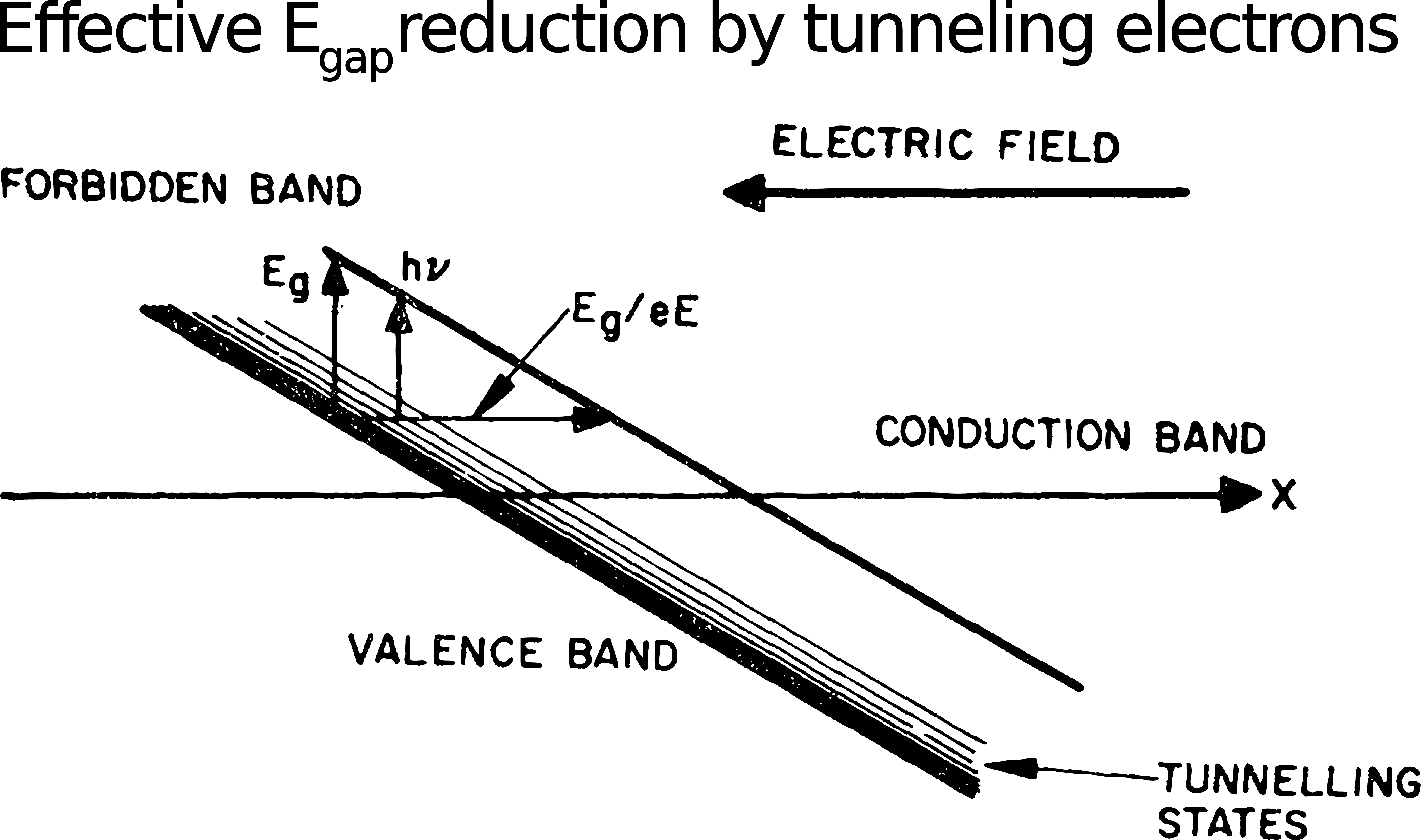}
       \caption{ }\label{subfig:Tunneling}
     \end{subfigure}\\[0.8\baselineskip]
     \begin{subfigure}[b]{\linewidth}
       \includegraphics[width=\linewidth]{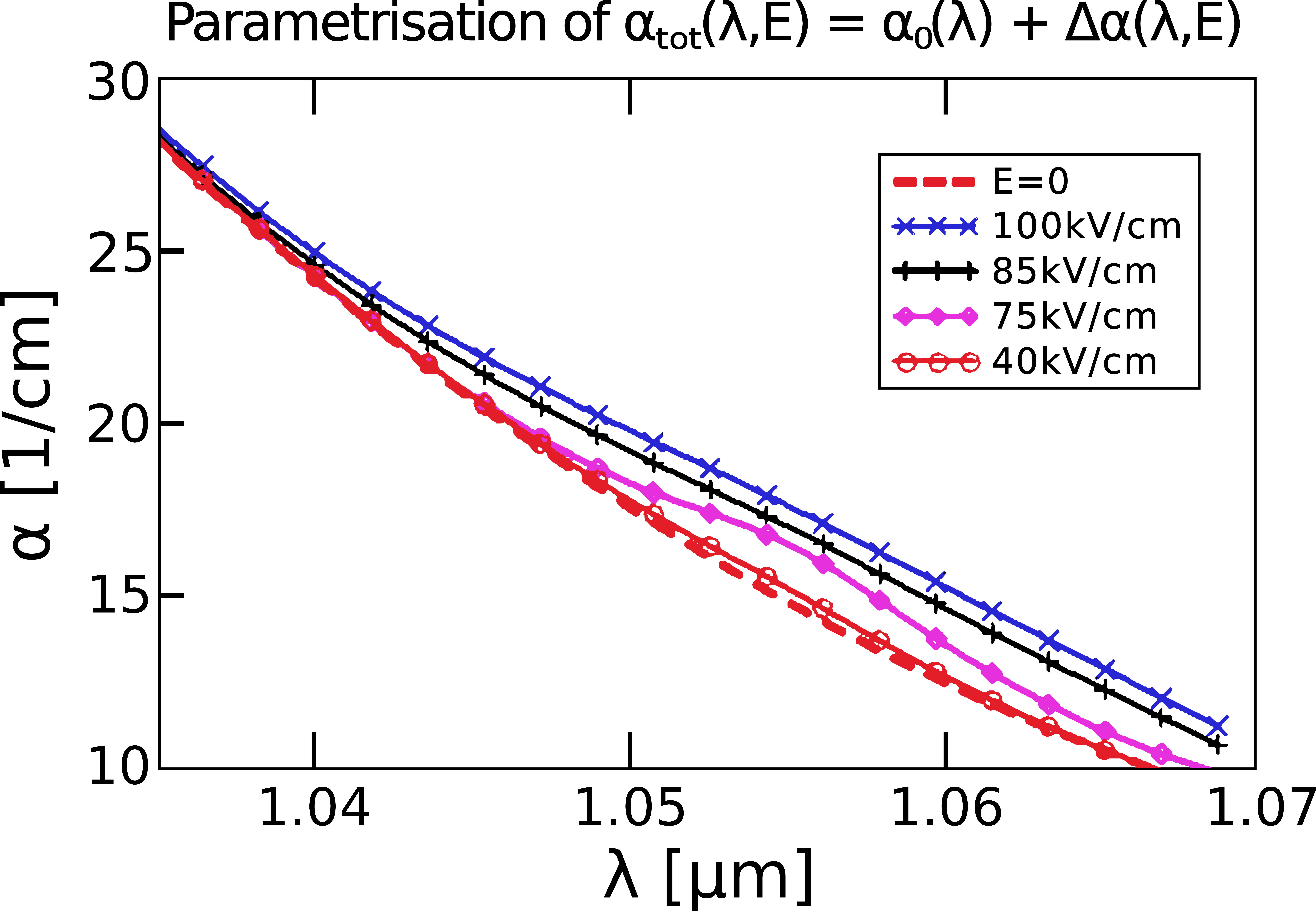}
       \caption{ }\label{subfig:Absorption}
     \end{subfigure}
   \end{minipage}
   \caption{Electric field dependence of the light absorption in silicon.
     (a) Field induced change of the absorption coefficient $\Delta \alpha (\lambda,E)$ (Fig.~3 of \cite{Wendland:1965}).
     (b) Physics explanation of $\Delta \alpha$ (Fig.~1 of \cite{Wendland:1965}).
     (c) Total absorption coefficient $\alpha (\lambda,E) = \alpha _0 (\lambda) + \Delta \alpha (\lambda,E)$, with $\alpha _0$ from Ref.~\cite{Green:2008} and $\Delta \alpha$ from Fig.~\ref{subfig:Wendland}.}
   \label{fig:Absorption}
 \end{figure}

 The intensity of light with photon energy $E_\gamma $ propagating in an absorbing homogeneous medium in the $z$~direction can be described by $I(z, E_\gamma) = I_0 \cdot e^{-\alpha (E_\gamma) \cdot z}$, with $I_0$ the intensity at $z = 0$ and the absorption coefficient $\alpha $.
 In Ref.~\cite{Wendland:1965} measurements of $\Delta \alpha(E_\gamma, E)$, the change of $\alpha(E_\gamma)$ with electric field, $E$, for $1.020\,\mathrm{eV} \leq E_\gamma \leq 1.220\,\mathrm{eV}$ and $40\,\mathrm{kV/cm} \leq E \leq 100\,\mathrm{kV/cm}$, temperatures between 72\,K and 300\,K for high-ohmic $n$- and $p$-type silicon are presented.
 Selected results at $23\,^\circ $C are shown in Fig.~\ref{subfig:Wendland}.
 It is found that $\Delta \alpha $ increases with $E$ with peaks at $E_\gamma = 1.059$\,eV and 1.175\,eV.
 The explanation of the increase in absorption, which was predicted in Refs.~\cite{Franz:1958, Keldysh:1958}, is given in Fig.~\ref{subfig:Tunneling}: The electric field causes tunneling states close to the band edges which result in a decrease of the effective band gap and therefore in an increase in the absorption of light with energy close to the band gap (1.124\,eV at 300\,K).
 As silicon has an indirect band gap, phonons are required to satisfy energy- and momentum conservation for the generation of electron--hole pairs by photons with energies close to the band gap.
 The peak at $E_\gamma = 1.059$\,eV is ascribed to phonon absorption and the one at $1.175\,$eV to phonon emission.
 At the temperatures of interest for the proposed measurements the latter dominates.

 From now on the wavelength $\lambda $[$\upmu$m]$~\approx 1.24 / E_\gamma $[eV] is used instead of $E_\gamma$, which is more common for optical measurements.
 Fig.~\ref{subfig:Absorption} shows the total absorption coefficient $\alpha _{tot} (\lambda, E) = \alpha _0 (\lambda ) +  \Delta \alpha(\lambda, E)$ at $23\,^\circ$C.
 The values of $\alpha _0$ are taken from Ref.~\cite{Green:2008} and the ones for $\Delta \alpha$ are obtained by digitizing the plot of Fig.~\ref{subfig:Wendland} and a cubic spline interpolation.

\begin{figure}[!ht]
  \centering
    \includegraphics[width=0.70\textwidth]{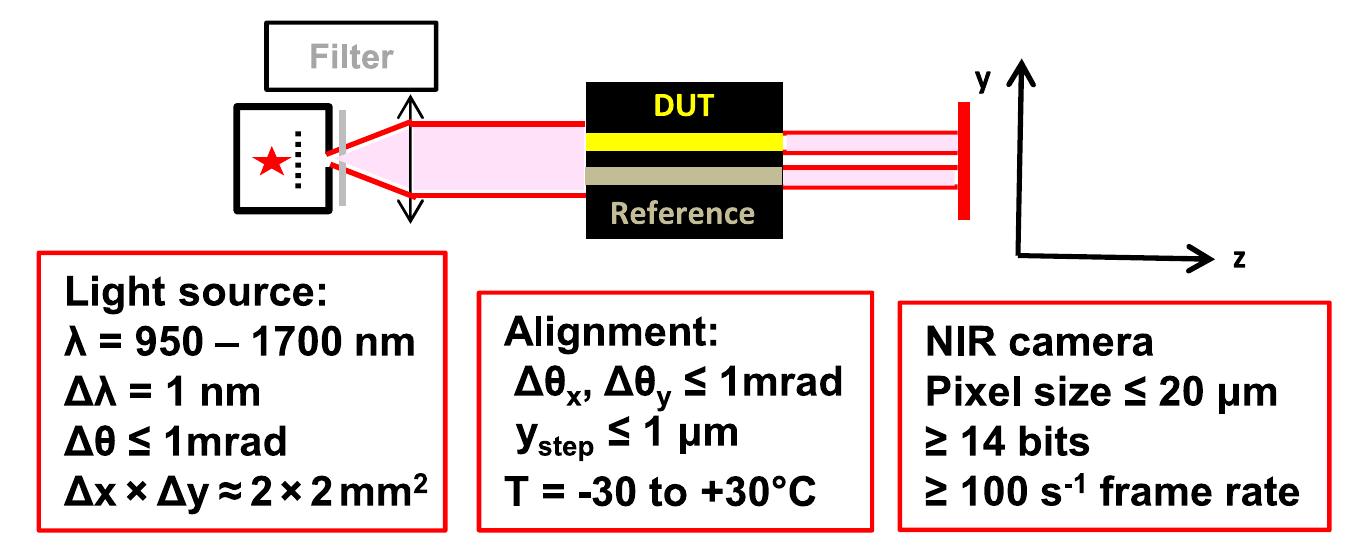}

   \caption{Schematic layout of the proposed set-up for the electro-optical measurement of the electric field in silicon sensors.}
  \label{fig:Layout}
 \end{figure}

 Fig.~\ref{fig:Layout} shows a schematic drawing of the proposed set-up.

 The specifications of the light beam at the DUT (Device under Test) are:
 a resolution $\Delta \lambda \approx 1$\,nm for wavelengths $900 < \lambda < 1350$\,nm,
 a beam spot of $2\,\mathrm{mm} \times 2\,\mathrm{mm}$,
 an angular dispersion of $ \lesssim 5$\,mrad, and
 a power density of 0.1\,mW/cm$^2$ for the selected $\Delta \lambda$~interval.
 In the wavelength interval, $1030\,\mathrm{nm} < \lambda < 1080\,\mathrm{nm}$ the spectral distribution should be smooth.
 In addition, a step-wise reduction of the light intensity by a factor of up to 1000 using gray filters is required.
 
 Detailed simulations show that the specifications can be essentially achieved by a tungsten-halogen lamp with a Czerny--Turner monochromator;
 however the small angular spread can be reached only in one dimension.
 All specifications can be achieved by a Supercontinuum Laser~\cite{Gunzow:2019} coupled to a Czerny--Turner monochromator.
 However, as primary laser the standard Yb~laser cannot be used, as it results in a strongly varying intensity spectrum in the wavelength interval of interest.

 The photon beam is directed onto an assembly of the DUT and the \emph{Reference}.
 Initially, for running-in the apparatus, a non-irradiated square pad diode of dimension $ a \times a \times d = 5\,\mathrm{mm} \times 5\,\mathrm{mm}\times 150\,\upmu $m polished on the entrance and exit sides (area $a \times d$) will be used as DUT.
 The direction of the beam is normal to the entrance window and precisely parallel to the surface of the DUT.
 The assembly can be moved vertically under remote control with sub-micron accuracy.
 For the precise alignment of the DUT onto the beam, it can also be rotated around the $x$- and $z$-axes.
 The intensity distribution for light with $\lambda \gtrsim 1.35\,\upmu$m, where silicon is practically transparent, can be used to check the alignment, the angular dispersion of the light beam and the quality of the DUT polishing.

 The \emph{Reference} serves as a monitor for the measurements: 
 With a polished SiO$_2$ slab the light flux and the alignment can be monitored, and a non-irradiated or an irradiated pad sensor at zero voltage serves as normalization for the voltage dependence of the absorption in the DUT.

 For the photo-detection a high speed NIR (near infrared) camera with an InGaAs CCD will be used.
 Cameras with $320 \times 256$ pixels with $20\,\upmu$m pitch, a quantum efficiency exceeding 50\% for wavelengths between 0.95 and 1.7\,$\upmu$m and 14\,bit readout resolution are commercially available.

 \section{Simulations}
  \label{sect:Simulations}

 In order to investigate the feasibility of the proposed method, the following  calculation has been performed.
 The $\Delta \alpha$\,plot of Fig.~\ref{fig:Absorption} has been digitized for the four $E$-values shown.
 To obtain $\Delta \alpha (\lambda, E)$, spline interpolations in $\lambda $ of the digitized $\Delta \alpha$-values for $E=40\,\mathrm{kV/cm}$ and $E=100\,\mathrm{kV/cm}$ and interpolations in $E$ using $c(\lambda) \cdot E^{b(\lambda)}$ have been used.
 Values for $b$ between 2 to 3 are found, which approximately agrees with the expectation of $b \approx 2$.
 Using the $\Delta \alpha $-values at 75\,kV/cm and 100\,kV/cm gives very different values for $c (\lambda )$ and $b(\lambda) $, however the results for the $E$\,dependence of the transmission are not too different.
 Precise data for $\Delta \alpha (\lambda, E)$, which are required for the analysis, will be obtained with the proposed set-up using non-irradiated pad sensors.
 It is noted that the parametrization $c(\lambda) \cdot E^{b(\lambda)}$ will have to be changed, as the peak of $\Delta \alpha $ shifts towards lower photon energies with increasing electric field.
 For the total absorption, $\alpha _{tot} (\lambda, E) = \alpha _0 (\lambda) + \Delta \alpha (\lambda, E)$, the parametrization of the zero-field absorption coefficient, $\alpha _0(\lambda)$ from Ref.~\cite{Green:2008} has been used.

 For the calculation of the transmission, the classical formulae have been used.
 The Fresnel formulae at normal incidence for the transmission $\mathit{Tra}$ and the reflection $\mathit{Ref}$ at the interface between  media of refractive index 1 and $n(\lambda)$
 \begin{equation}\label{equ:Fresnel}
   \mathit{Tra}(\lambda) = 4 \cdot n(\lambda) / (n(\lambda ) +1)^2 \hspace{5mm} \mathrm{and} \hspace{5mm} \mathit{Ref}(\lambda) = 1 - \mathit{Tra}(\lambda ),
 \end{equation}
 and for the transmission of a slab of thickness $a$ and absorption coefficient $\alpha(\lambda )$
 \begin{equation}\label{equ:Trans}
   \mathit{Tr}(\lambda, \alpha, a) = \frac{ \mathit{Tra}(\lambda)^2 \cdot e^{- \alpha (\lambda) \cdot a} } { 1 -(\mathit{Ref} (\lambda) \cdot e^{- \alpha (\lambda) \cdot a} )^2 },
 \end{equation}
 where the second term in the denominator takes into account the increase in transmission from multiple reflections in the slab.
 For $n(\lambda )$ the data from Ref.~\cite{Green:2008} have been used.

\begin{figure}[!ht]
   \centering
   \begin{subfigure}[a]{0.7\textwidth}
    \includegraphics[width=\textwidth]{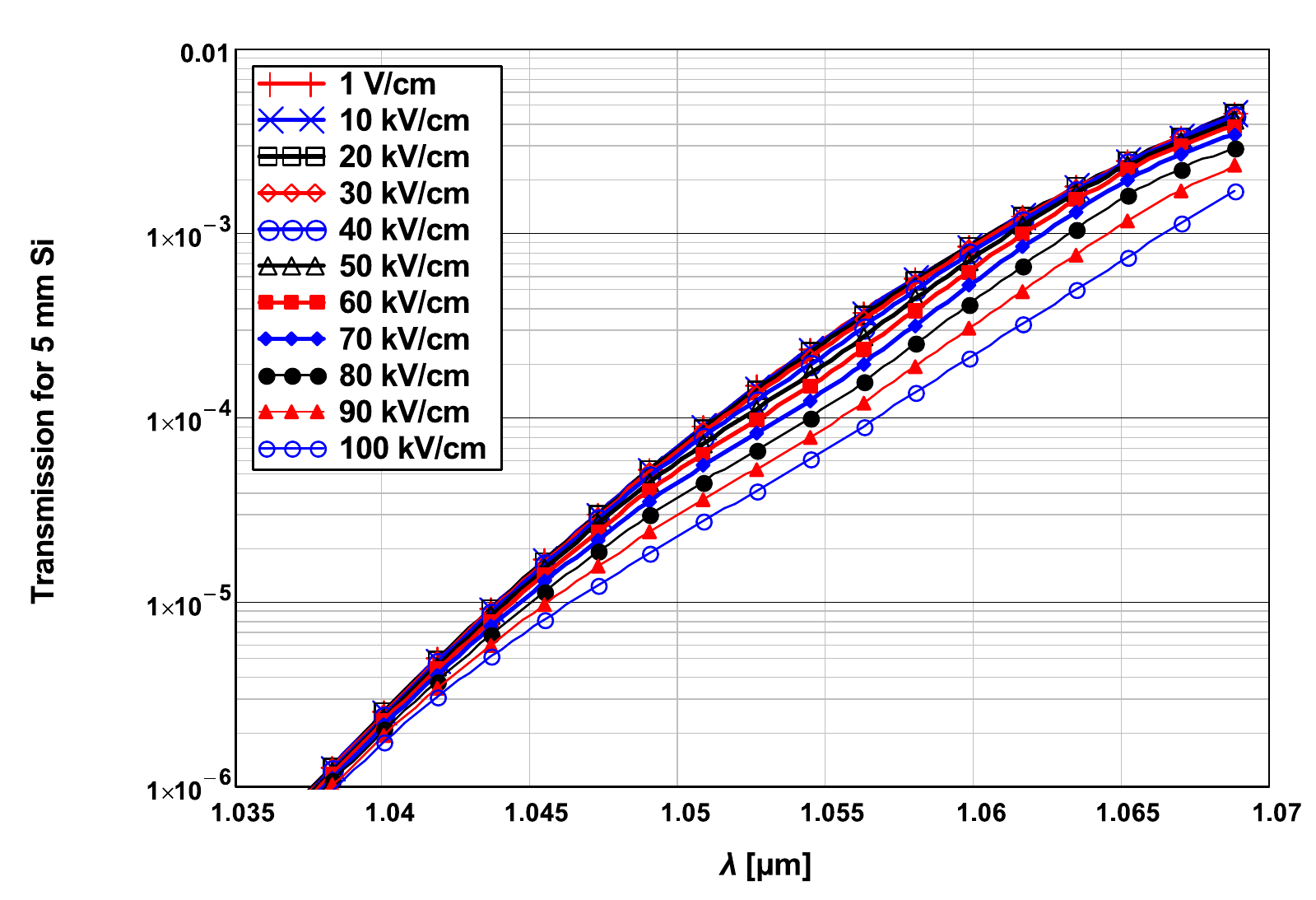}
     \caption{ }
    \label{fig:Tr5mm}
   \end{subfigure}%
    ~
\newline
   \begin{subfigure}[a]{0.65\textwidth}
    \includegraphics[width=\textwidth]{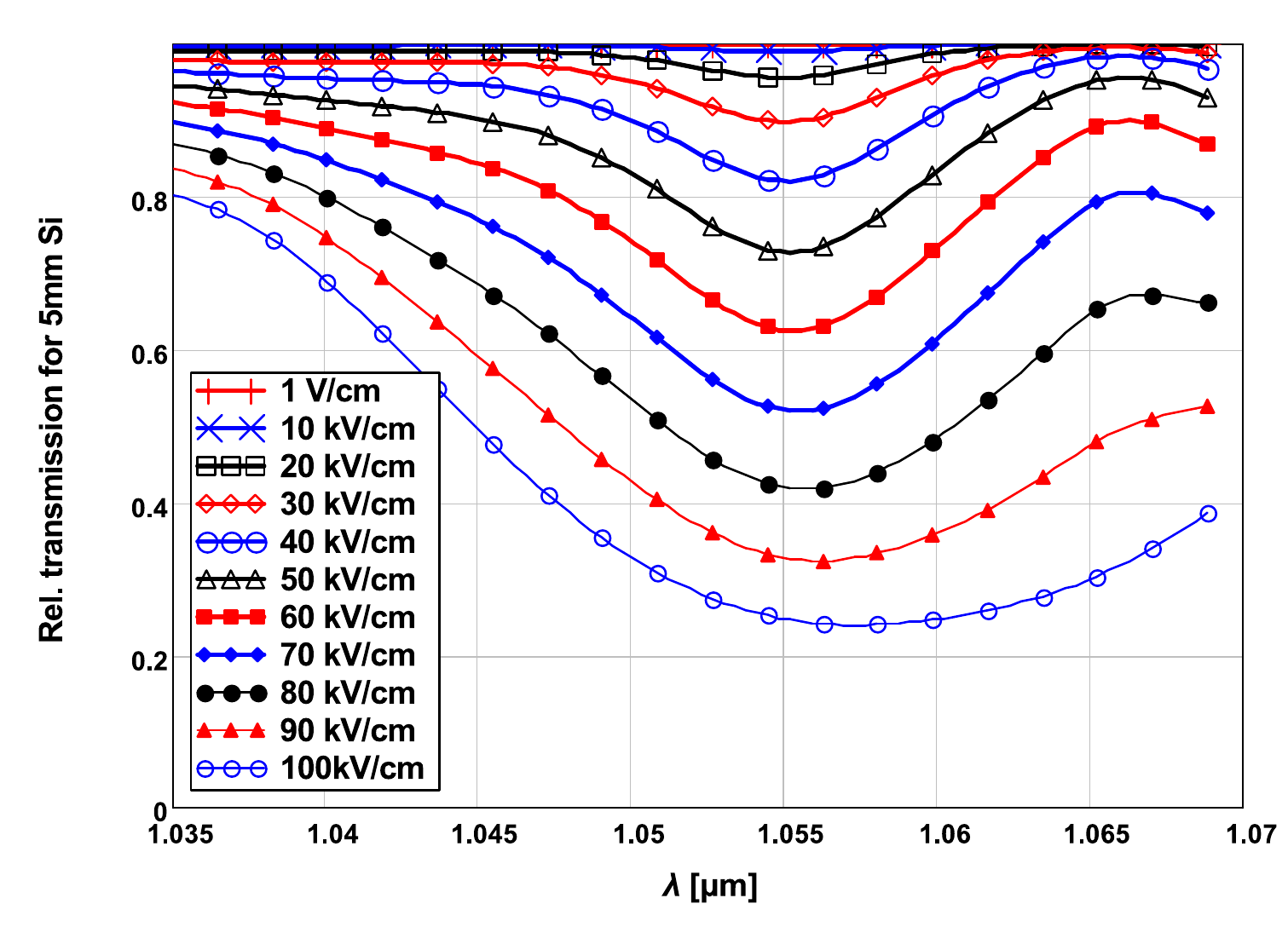}
     \caption{ }
    \label{fig:RelTr5mm}
   \end{subfigure}%
   \caption{ Simulated transmission, $\mathit{Tr}(\lambda , E)$, for 5\,mm silicon as a function of wavelength, $\lambda$, for selected values of the electric field, $E$.
    (a) Absolute transmission, and
    (b) transmission normalised to the value for $E = 0$. }
  \label{fig:Tr}
 \end{figure}

\begin{figure}[!ht]
  \centering
    \includegraphics[width=0.5\textwidth]{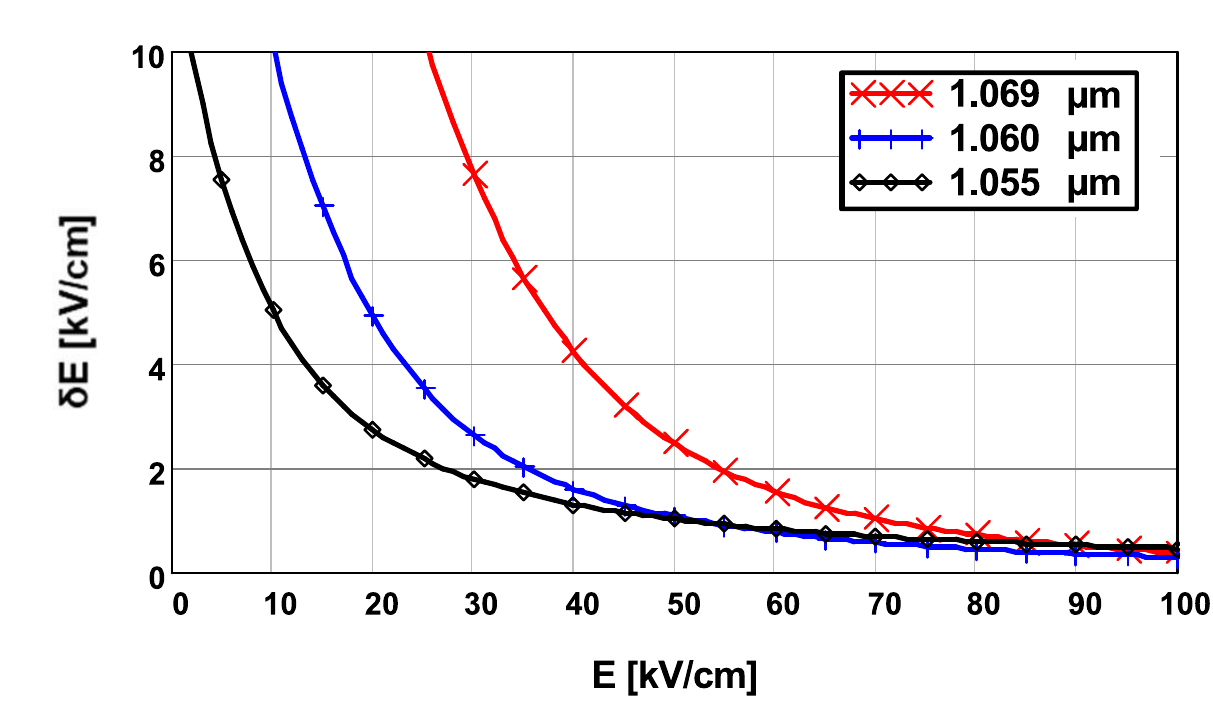}
   \caption{ Uncertainty, $\delta E$, of the determination of $E$ for selected values of $\lambda $ obtained from the ratio $\mathit{Tr}(\lambda , E)/\mathit{Tr}(\lambda , E = 0)$ assuming uncorrelated uncertainties $\delta \mathit{Tr}(\lambda ,E) = \delta \mathit{Tr}(\lambda ,E=0) = 1\%$.}
  \label{fig:dE}
 \end{figure}

 Fig.~\ref{fig:Tr5mm} shows the absolute transmission $\mathit{Tr}(\lambda , E, a)$ calculated for $a = 5$\,mm of silicon, for wavelengths between $1.035\,\upmu \mathrm{m}$ and $1.070\,\upmu \mathrm{m}$, and electric fields between 0 and 100\,kV/cm, and Fig.~\ref{fig:RelTr5mm} the relative transmission $ \mathit{Tr}_\mathrm{rel}(\lambda , E) = \mathit{Tr}(\lambda , E)/\mathit{Tr}(\lambda , E = 0) $).
 To estimate $\delta E$, the accuracy of the $E$\,determination, $E$ is reconstructed from $\mathit{Tr}_\mathrm{rel}(\lambda, E)$ assuming an uncorrelated uncertainty of 1\% for the two $\mathit{Tr}$-values.
 The results for $\delta E(E)$ for three $\lambda $-values are shown in Fig.~\ref{fig:dE}.
 For $E>20\,\mathrm{kV/cm}$ $\delta E$-values below 4\,kV/cm are found.
 The high accuracy of the determination of $E$ from the relative transmission measurement may appear surprising.
 It can be understood from Eq.~\ref{equ:Trans}:
 For the conditions of the measurements the term $(\mathit{Ref} \cdot e^{- \alpha \cdot a}) ^2$ in the denominator is below $10^-4$ and can be ignored relative to 1. As a result $\mathit{Tr}_\mathrm{rel} (\lambda , E) = e^{- \Delta \alpha (\lambda , E) \cdot a}$ does not depend on the absolute $\mathit{Tr}$-values.
 In addition, many systematic uncertainties cancel when $E$ is determined from $\mathit{Tr}_\mathrm{rel}$.

\begin{figure}[!ht]
   \centering
   \begin{subfigure}[a]{0.5\textwidth}
    \includegraphics[width=\textwidth]{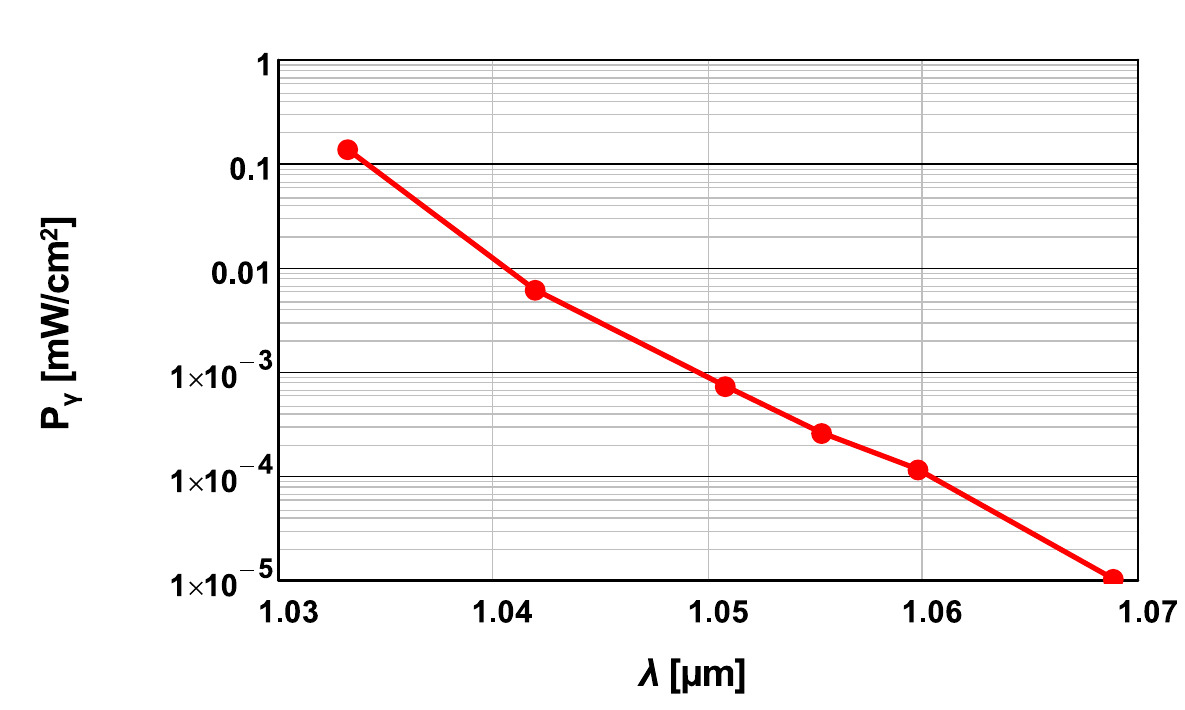}
     \caption{ }
    \label{fig:Pgamma}
   \end{subfigure}%
    ~
   \begin{subfigure}[a]{0.5\textwidth}
    \includegraphics[width=\textwidth]{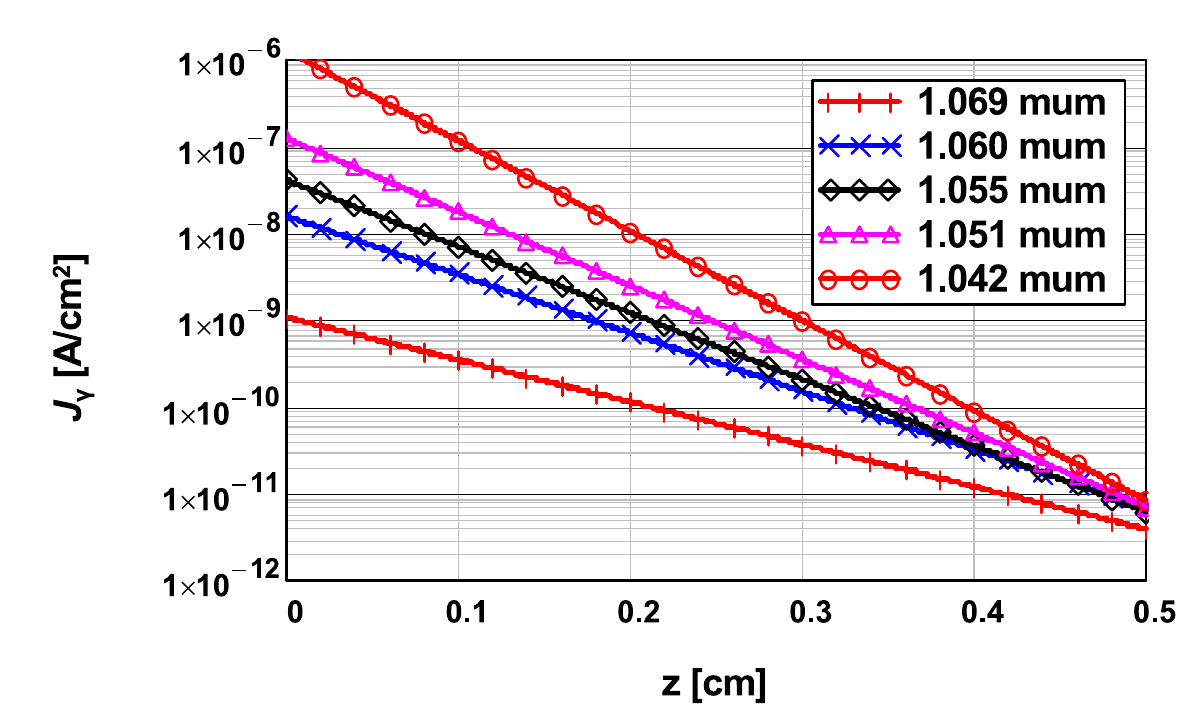}
     \caption{ }
    \label{fig:Jgamma}
   \end{subfigure}%
   \caption{
    (a) Required light power density, $P_\gamma (\lambda)$, and
    (b) resulting position-dependent photo-current density, $J_\gamma (z,\lambda) $,
        for the transmission measurement with an accuracy of 1\% in 1~second for a 5\,mm light path at an electric field of 100\,kV/cm. }
  \label{fig:PJ}
 \end{figure}

  From the simulations the required photon flux, $\mathit{Fl}_\gamma (\lambda)$, and the current density, $J_\gamma (z, \lambda)$, from the photons converting in the detector are estimated.
  For a statistical uncertainty $\delta \mathit{Tr} $, a measurement time $t_{meas}$, a photo-detector area $A = \Delta x \times \Delta y$ for the measurement of the position dependence of the electric field, and a photon-detection efficiency, $\mathit{pde}(\lambda )$, the required photon flux is:
 \begin{equation}\label{equ:Fl}
   \mathit{Fl}_\gamma (\lambda) = \big(\delta \mathit{Tr}^2 \cdot \mathit{pde}(\lambda) \cdot A \cdot \mathit{Tr}(\lambda , E_{max}, a) \cdot t_{meas}\big)^{-1},
 \end{equation}
 with $E_{max} = 100\,\mathrm{kV/cm}$, the maximum field for the planned measurements.
 The corresponding power density for the selected $\Delta \lambda $ interval of the measurement is $P_\gamma (\lambda) = \mathit{Fl}_\gamma (\lambda) \cdot E_ \gamma (\lambda) $.
 Fig.~\ref{fig:Pgamma} shows $P_\gamma (\lambda)$ for
 $ \delta \mathit{Tr} = 1\,\%$,
 $ \mathit{pde}(\lambda) = 50\,\%$,
 $ \Delta x \times \Delta y = 1\,\mathrm{mm} \times 20\,\upmu$m, and
 $ t_{meas} = 1$\,s.
 The value for $\Delta y$ has been chosen not to significantly worsen the measurement of the electric field: the change of $E$ over $\Delta y = 20\,\upmu$m for a non-irradiated diode with a doping of $10^{13}\,\mathrm{cm}^{-3}$ is $\Delta E = \mathrm{d}E/\mathrm{d}y \cdot \Delta y \approx 3$\,kV/cm.
 The choice of the value of $\Delta x$ is adequate for determining the electric field in a pad detector, which does not depend on $x$, but has to be reduced for strip- or pixel detectors.
 The strong dependence of $P_\gamma $ on $\lambda $ necessitates the use of filters, as shown in Fig.~\ref{fig:Layout}, and/or different choices of $t_{meas}$ to adapt to the dynamic range of the photo-detector.

 For estimating the photo-current density, $\mathit{Fl}_\gamma (\lambda) $ has to be multiplied with $\mathit{Tra} (\lambda )$ (Eq.~\ref{equ:Fresnel}) to obtain the photon flux entering the silicon at $z = 0$.
 The $z$-dependent current density is approximately
  \begin{equation}\label{eq:Jgamma}
    J_\gamma (z, \lambda ) = q_0 \cdot Fl_\gamma (\lambda) \cdot Tra (\lambda ) \cdot  \int _0 ^d \Big(\alpha(\lambda, E(y))\cdot e^{- \alpha (\lambda, E(y)) \cdot z} \Big) ~ \mathrm{d}y,
  \end{equation}
 with $q_0$ the elementary charge.
 Fig.~\ref{fig:Jgamma} shows $J_\gamma(z,\lambda)$ at $E=100\,\mathrm{kV/cm}$ for several values of $\lambda $ for the photon flux required to achieve a 1\,\% measurement accuracy for 1~second measurements, which can be compared to the measured dark current of irradiated diodes.
 For a $200 \,\upmu$m thick $n^+p$ diode irradiated by 24\,GeV/c protons to a fluence of $1.3 \times 10^{16}\,\mathrm{cm}^{-2}$ after annealing for 80' at $60\,^\circ $C, Ref.\,\cite{Schwandt:2018} reports a value of $3 \times 10^{-4}\,\mathrm{A}/\mathrm{cm}^{2}$ at $-20\,^\circ $C and 1000\,V.
 The value at $-30\,^\circ$C is $10^{-4}\,\mathrm{A}/\mathrm{cm}^{2}$.
 As these values are significantly larger than the current densities from the photons, no significant change of the electric field in the diode is expected.

 It should be noted that given the quite large uncertainties of $\Delta \alpha (\lambda, E)$ the results can be considered only an estimate.
 However, they indicate that a precise determination of the electric field in silicon sensors with the proposed method appears feasible.

  \section{Summary and Conclusions}
   \label{sect:Conclusions}

 A  method for measuring the electric field in silicon sensors by electro-optical imaging is proposed.
 It makes use of the Franz--Keldysh effect, the increase with electric field of the absorption coefficient of photons with energies close to the silicon band gap.
 A simulation using published data shows that for an edge-on illumination of silicon sensors an accurate field measurement appears to be possible: The estimated uncertainty is in the range of 1~to 4\,kV/cm for fields exceeding 20\,kV/cm and a 5\,mm light path.

 A schematic set-up is presented and possible choices for its components are discussed.
 After running-in the set-up with non-irradiated pad sensors and a precise measurement of the Franz--Keldysh effect, it is intended to determine the electric fields in different radiation-damaged pad and segmented sensors.
 Although the knowledge of the electric field in radiation-damaged silicon sensors is important for their understanding and optimization in the harsh environment of collider experiments, no satisfactory method for its determination exists so far; the proposed experiment may be such a method.
 The experiment will also provide valuable data on the change of the light absorption as a function of irradiation.
 It should be noted that silicon becomes birefringent at high electric fields~\cite{Gutkin:1972}, which may possibly be used to determine, in addition to the absolute value of the electric field, its direction.

 First steps towards the realization of a set-up for the electro-optical imaging of electric fields in silicon detectors are underway at the Detector Lab of Hamburg University.
 In addition, the work on more detailed simulations and analysis methods, which include the angular spread of the photon beam and the effects of diffraction, has started.

\section*{Acknowledgments}
The  work  of A.V. is supported by the Deutsche Forschungsgemeinschaft (DFG, German Research Foundation) under Germany's Excellence Strategy -- EXC 2121 Quantum Universe - 390833306.

  \section*{Bibliography}
   \label{sect:Bibliography}

\end{document}